\documentclass[floats,floatfix,showpacs,amssymb,prd,twocolumn,superscriptaddress,nofootinbib,nolongbibliography,reprint]{revtex4-2}

\usepackage{amssymb,amsmath,verbatim,mathtools,needspace,enumitem,etoolbox,graphicx,physics,microtype,afterpage,xspace,tabularx,lmodern,multirow}
\usepackage{gensymb}
\usepackage[dvipsnames, usenames]{xcolor}
\definecolor{linkcolor}{rgb}{0.0,0.3,0.5}
\usepackage[unicode, colorlinks=true, linkcolor=linkcolor, citecolor=linkcolor, filecolor=linkcolor, urlcolor=linkcolor, linktocpage, breaklinks]{hyperref}
\usepackage[all]{hypcap}
\usepackage[T1]{fontenc}
\usepackage[utf8]{inputenc}
\usepackage[usenames,dvipsnames]{xcolor}
\hypersetup{colorlinks=true,citecolor=romared,linkcolor=romared,urlcolor=romared}

\definecolor{romared}{RGB}{142,0,28}

\newcommand{\be}{\begin{equation}}
\newcommand{\ee}{\end{equation}}

\def\be{\begin{equation}}
\def\ee{\end{equation}}
\newcommand{\beq}{\begin{eqnarray}}
\newcommand{\eeq}{\end{eqnarray}}

\usepackage{aas_macros}
\usepackage{makecell}
\usepackage{soul}

\renewcommand{\d}{{\rm d}}

  \newcommand{\aIII}{{a}_{\rm III}}
  \newcommand{\bIII}{{b}_{\rm III}}
  \newcommand{\zIII}{{z}_{\rm III}}

\newcommand{\software}[1]{\texttt{#1}}
\newcolumntype{Y}{>{\centering\arraybackslash}X}

\newcommand{\crate}{\ensuremath{^{12}\rm{C}\left(\alpha,\gamma\right)^{16}\!\rm{O}}}

\newcommand{\jhu}{\affiliation{Department of Physics and Astronomy, Johns Hopkins University, 3400 North Charles
Street, Baltimore, Maryland 21218, USA}}

\begin{document}
\title{Beyond the far side: Observing black hole mergers
beyond the pair-instability mass gap \\ with next-generation gravitational wave detectors}

\begin{abstract}
Stellar evolution predicts the existence of a mass gap for black hole remnants produced by pair-instability supernova dynamics, whose lower and upper edges are very uncertain. 
We study the possibility of constraining the  location of the upper end of the pair-instability mass gap, which is believed to appear around ${m_\text{\tiny min}} \sim130M_\odot$, using gravitational wave observations of compact binary mergers with next-generation ground-based detectors.
While high metallicity may not allow for the formation of first-generation black holes on the ``far side'' beyond the gap, metal-poor environments containing Population III stars could lead to such heavy black hole mergers.
We show that, even in the presence of contamination from other merger channels, next-generation detectors will measure the location of the upper end of the mass gap with a relative precision close to $\Delta {m_\text{\tiny min}}/{m_\text{\tiny min}} \simeq 4\% (N_\text{\tiny det}/100 )^{-1/2}$ at 90\% C.L., where $N_\text{\tiny det} $ is the number of detected mergers with both members beyond the gap. 
These future observations could reduce current uncertainties in nuclear and astrophysical processes controlling the location of the gap.
\end{abstract}

\author{Gabriele Franciolini}
\email{gabriele.franciolini@cern.ch}
\affiliation{CERN, Theoretical Physics Department,
Esplanade des Particules 1, Geneva 1211, Switzerland}

\author{Konstantinos Kritos}
\email{kkritos1@jhu.edu}
\jhu

\author{Luca Reali}
\email{lreali1@jhu.edu}
\jhu

\author{Floor Broekgaarden}
\email{fsb2127@columbia.edu}
\jhu
\affiliation{Department of Astronomy and Columbia Astrophysics Laboratory,
Columbia University, 550 W 120th St, New York, NY 10027, USA}
\affiliation{Simons Society of Fellows, Simons Foundation, New York, NY 10010, USA}

\author{Emanuele Berti}
\email{berti@jhu.edu}
\jhu

\hfill {\footnotesize ET-0006A-24. CERN-TH-2024-003}

\date{\today}
\maketitle

\section{Introduction}
\label{Sec:Introduction}

Gravitational waves (GWs) are a new probe into the mass spectrum of black holes (BHs). 
The LIGO-Virgo-KAGRA collaboration detected GW transient events produced by mergers of compact objects at cosmological distances, with a majority of these mergers coming from binary BHs (BBHs)~\cite{LIGOScientific:2018mvr,LIGOScientific:2020ibl,LIGOScientific:2021djp,Nitz:2021zwj,Mehta:2023zlk,Olsen:2022pin}.
Some of the events in these catalogs come from the merger of BHs with masses larger than those routinely observed in X-ray binary systems: for example, the components of GW150914, the first GW observation of a BH-BH merger, have masses of the order of $\sim30\,M_\odot$~\cite{LIGOScientific:2016aoc}. 

Theory predicts an upper mass gap for the BH mass distribution (to be distinguished from the hypothetical ``lower mass gap'' between the heaviest neutron stars and the lightest BHs) in the range $45-130\,M_{\odot}$, where a dearth of BHs is theoretically expected due to the pair-instability supernova (PISN) mechanism~\cite{1964ApJS....9..201F,1967PhRvL..18..379B,Woosley:2021xba}.  Stars with He core masses in the range from $\simeq 64M_\odot$ to $\simeq130M_\odot$, that would form such BH masses, produce energetic gamma photons which create electron-positron pairs after the interaction with an atomic nucleus in the star.
 The consequent reduction of radiation pressure in the stellar interior causes its implosion, and the temperature rising to a few$\times10^9$K results in the explosive burning of oxygen, which destroys the star without leaving a compact remnant behind~\cite{1964ApJS....9..201F,1967ApJ...148..803R,1967PhRvL..18..379B,1983A&A...119...61O,Woosley:2021xba}.
When the He core of a star is more massive than $\simeq130M_\odot$, its gravitational potential is believed to be large enough for it to evade the destructive explosion, and thus directly collapse into a massive BH above the upper mass gap~\cite{1984ApJ...280..825B,Heger:2001cd}.
The exact boundaries of the upper mass gap are unknown and they are ultimately metallicity-dependent, but the dominant uncertainty comes from the $^{12}$C($\alpha$,$\gamma$)$^{16}$O nuclear reaction rate~\cite{Belczynski:2016jno,deBoer:2017ldl,Farmer:2019jed,2020ApJ...902L..36F,2021MNRAS.501.4514C,2021ApJ...912L..31W,Mehta:2021fgz,Farag:2022jcc,Shen:2023rco}.

On the observational side, the LIGO-Virgo-KAGRA collaboration found that the merger rate shows a decline as a function of the primary component mass which may be associated with the existence of the upper mass gap~\cite{LIGOScientific:2021psn,Baxter:2021swn}, but the evidence from current data is still inconclusive. 
The LIGO-Virgo-KAGRA collaboration also concluded that one of the components of the GW190521 merger event is confidently within the upper mass gap. Follow-up work has suggested that GW190521 could also straddle the mass gap with one BH above the mass gap~\cite{Edelman:2021fik,Fishbach:2020qag,Nitz:2020mga}, and Ref.~\cite{Wadekar:2023gea} recently reported candidate GW events above the mass gap. 
Modelling of the GW observations consistently finds a feature near the lower end of the upper mass gap (i.e., a change in power law slope or a Gaussian peak) around $\sim 35 M_{\odot}$. This feature seems robust (see e.g.~\cite{LIGOScientific:2021psn,Farah:2023swu,Sadiq:2021fin,Callister:2023tgi}) and it has been linked to pulsational pair instability supernova (PPISN)~\cite{Talbot:2018cva,Stevenson:2019rcw,Belczynski:2020bca,Karathanasis:2022rtr}, but see~\cite{Antonini:2022vib,Briel:2022cfl,Li:2022ApJ,Hendriks:2023yrw} for alternative possibilities.

Several proposed formation channels can produce BHs above the upper mass gap.
For example, BHs above the upper mass gap could be formed from very massive stars born in very low metallicity environments ($Z\lesssim5\%Z_\odot$) due to the reduction of strong stellar winds~\cite{Spera:2017fyx}.  Here $Z_\odot\simeq1.4\%$ is the solar metallicity, i.e., the mass fraction of metals (elements heavier than H and He) in the Sun~\cite{2009ARA&A..47..481A}.
This possibility can be realized in Population III (henceforth Pop III) stars, the first low-metallicity stars at very high redshift~\cite{Bromm:2003vv,Tanikawa:2021qqi}.
For example, Reference~\cite{Hijikawa:2021hrf} showed that a fraction of these BBHs would have component masses beyond the upper edge of the upper mass gap, which we denote here by $m_\text{\tiny min}$, following the notation of~\cite{Spera:2022byb}.
Such BBHs tend to merge at higher redshift ($z>5$)~\cite{Hijikawa:2021hrf}, and as such they are out of reach for current ground-based GW observatories, but they may be detectable by next-generation GW observatories~\cite{Ng:2020qpk,Ezquiaga:2020tns}.
Other popular astrophysical contamination mechanisms include hierarchical BH mergers in the cores of stellar clusters~\cite{Gerosa:2017kvu,Fishbach:2017dwv,Rodriguez:2019huv,Gerosa:2021mno}, primordial black holes formed by the collapse of large overdensities shortly after the Big Bang~\cite{Bird:2016dcv, Sasaki:2016jop,Clesse:2016vqa, Ali-Haimoud:2017rtz,Clesse:2020ghq,DeLuca:2020sae,Kritos:2020wcl}, remnants of runaway stellar mergers in crowded systems~\cite{Kremer:2020wtp,DiCarlo:2019fcq}, or the core collapse of rapidly rotating massive stars from progenitors with He cores $\gtrsim 130\,M_\odot$ (``collapsars''), which could contaminate the mass gap ``from above'' while leading to long GRBs, r-process nucleosynthesis, and GWs of frequency $\sim 0.1-50$\,Hz from nonaxisymmetric instabilities~\cite{Siegel:2021ptt}.
Any contamination of the upper mass gap may complicate the measurement of the gap edges and potentially jeopardize its determination, even when accounting for the large number of observations expected with next-generation detectors. 
In the most motivated cases, however, the contamination is expected to be more relevant in the lighter-mass portion of the gap (as predicted e.g. by hierarchical merger models), while $m_\text{\tiny min}$ should provide a cleaner signature of the presence of a gap.

Earlier work has investigated the detectability of BHs above the mass gap. Following a model-independent approach, Reference~\cite{Mangiagli:2019sxg} showed that the detection rate of isolated BBHs assembled in the galactic field having at least one BH component above the upper mass gap could be in the range $10$--$460\,\rm yr^{-1}$ (or even higher) with a next-generation GW observatory at signal-to-noise (SNR) threshold of 12. 
This broad rate estimate is strongly dependent on binary-star physics, in particular on the details of the cosmic star formation rate and the metallicity prescription~\cite{Santoliquido:2023}, but it is consistent with the detection rate of $\approx 126\,\rm yr^{-1}$ reported in Ref.~\cite{Hijikawa:2021hrf}.
The BBH merger rate density evolution and mass spectrum from these metal-free stars are also strongly dependent on the star formation rate, and such mergers could be rare~\cite{Santoliquido:2023}.

In this work, we investigate the {\it detectability} and {\it measurability} of the merger of BHs originating from Pop III stars with next-generation GW detectors, such as the Einstein Telescope (ET) and Cosmic Explorer (CE).
Our analysis relies on Monte Carlo injection of simulated merger events and on hierarchical Bayesian inference to extract the key parameters of the mass function of heavy BHs beyond the gap, including binary parameter estimation uncertainties.
As we focus on BHs beyond the upper edge of the upper mass gap, we only consider BBHs with both component masses above $m_\text{\tiny min}$.
Given the aforementioned uncertainties on the detection rate of BBHs with a primary mass larger than $m_\text{\tiny min}$, we treat the rate normalization as a free parameter and we make different assumptions on its value.
The rest of this work is structured as follows. In Sec.~\ref{Sec:Methods} we describe our assumptions about the BBH population model used to perform the analysis.
In Sec.~\ref{Sec:Results} we show our main results, while in Sec.~\ref{sec:astroimpl} we discuss their astrophysical implications and present our conclusions.

\section{Population models}
\label{Sec:Population_model}

Remnant BHs on the far side of the mass gap, with masses larger than $m_{\rm \tiny min}$, may be produced from the evolution of very massive stars with metallicities $Z<5\% Z_\odot$.
These massive stars are produced from the collapse of pristine gas clouds at high redshift, when the interstellar medium is only poorly enriched with metals.
We focus on the first (Pop III) stars, which form in mini-halos. A large fraction of these stars have negligible multiplicities, and roughly half of all Pop III stars are expected to form in binaries~\cite{Susa:2014moa}.
These Pop III stars (formed from the metal-free baryonic gas in the early Universe) can produce heavy BHs, because very massive stars can directly collapse into BHs with mass greater than $m_{\rm\tiny min}$. Therefore, binaries of Pop III stars can give rise to BBHs with at least one such massive BH, according to population synthesis studies (see e.g.~\cite{Hijikawa:2021hrf}).
The resulting mass function of primary and secondary BHs depends on the progenitor masses and on the complex binary evolutionary process.

\begin{figure*}
\centering
\includegraphics[width=0.314 \textwidth]{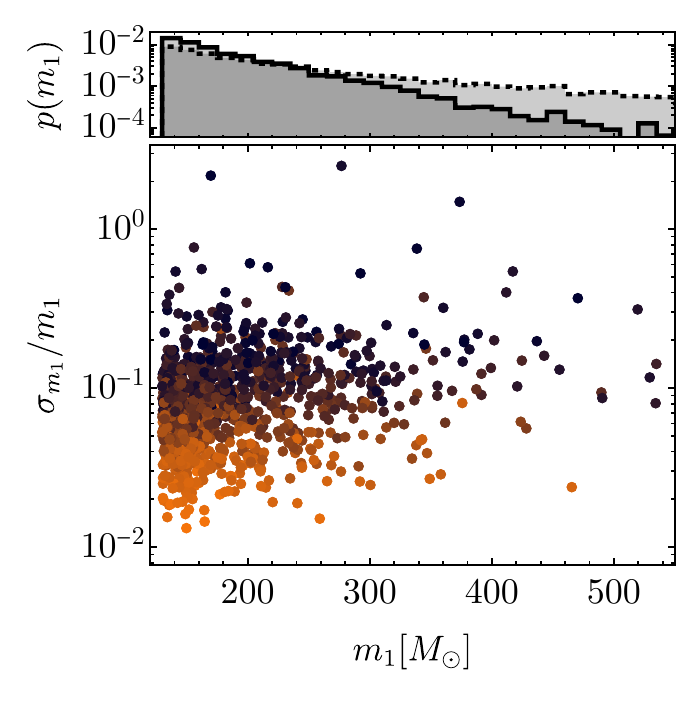}
\includegraphics[width=0.314 \textwidth]{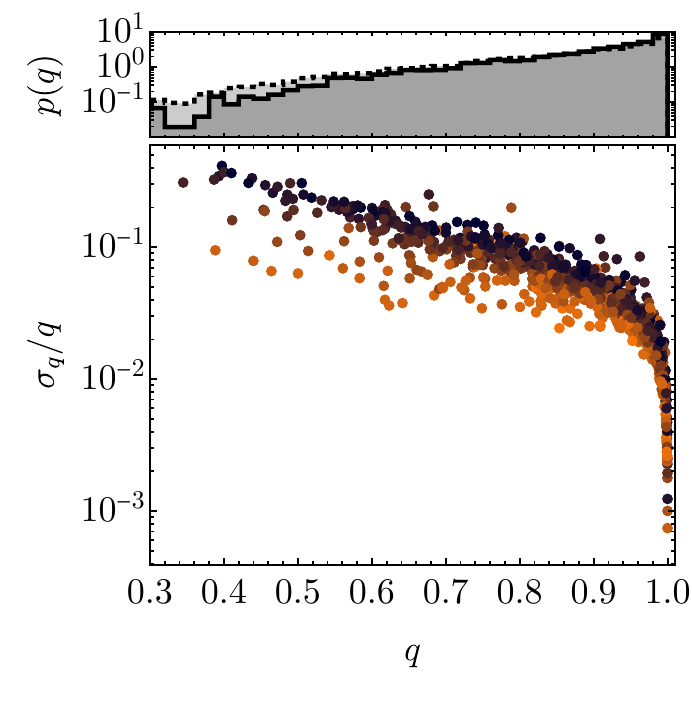}
\includegraphics[width=0.357
 \textwidth]{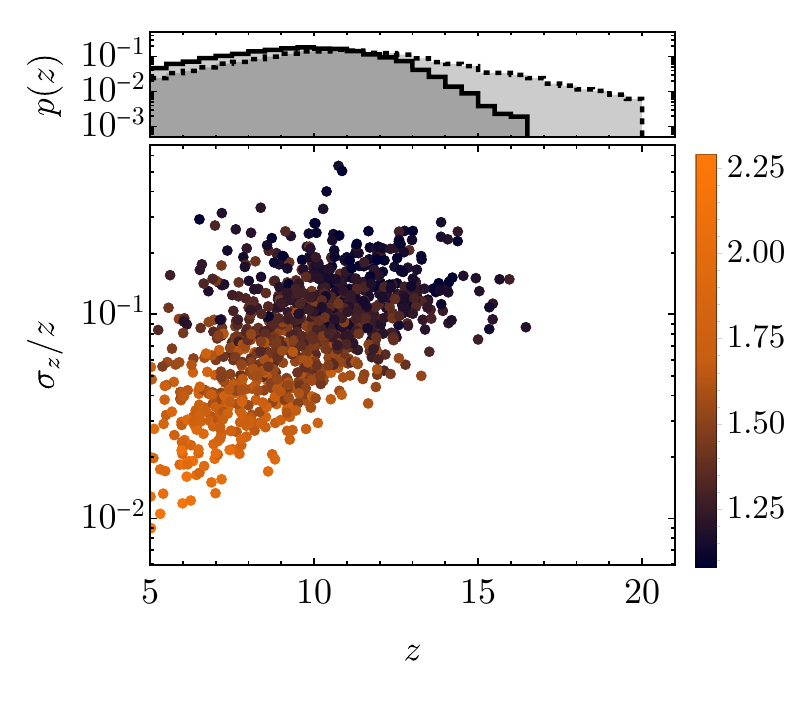}
\caption{
Top panels:
Distribution of primary mass (left), mass ratio (center), and redshift (right) for the injected Pop III binary channel.
The black dashed and black continuous histograms refer to the intrinsic and observed populations, respectively.
Bottom panels:
Parameter estimation $1\sigma$ uncertainty on the corresponding observable for each event in the synthetic catalog.
Each point adopts the color scheme indicated on the right side, reporting the color code for $\log_{10}({\rm SNR})$.
There is a positive correlation between high SNRs (light colors) and low redshifts, and also between PE accuracy and SNR.
}
\label{fig:pop_errors}
\end{figure*}

Motivated by Refs.~\cite{Hijikawa:2021hrf,Costa:2023xsz}, we assume the primary mass to be distributed according to a power law with spectral index $-\alpha<0$.
We choose a fiducial value of $130M_\odot$ for $m_{\rm\tiny min}$.
For simplicity, we do not introduce a window term below $m_{\rm\tiny min}$, as typically adopted in LIGO-Virgo-KAGRA analyses to smooth out the sharp cutoff.
We also assume a maximum BH mass of $m_\text{\tiny max}=600M_\odot$, but our results are not strongly sensitive to the choice of $m_{\rm\tiny max}$ for two reasons.
The first reason is that the BH mass distribution drops sharply following the power-law mass distribution beyond $\approx 130M_\odot$: the most massive BHs are rare, due to the rarity of the massive stars that produce them. 
The second reason is that BBHs from Pop III stars merge at high redshift, and that GW detectors observe redshifted masses.
It is expected that next-generation detectors will be sensitive only above $\sim {\rm few}\,\rm Hz$, while BBHs with a total mass of a hundred solar masses that merge at $z\sim10$ would be redshifted to merger frequencies in the sub-Hz regime.
Those heavy events would therefore be undetectable even by next-generation detectors.

Having drawn the primary mass $m_1$ from this power law distribution, we compute the mass of the secondary component $m_2=q\, m_1$ by sampling the mass ratio $q$ from another power law with index $\beta$, normalized in the range $[q_\text{\tiny min}=m_{\rm\tiny min}/m_1,\,1]$.
The lower value in $q$ arises from the minimum allowed BH mass in our model.

Following Ref.~\cite{Hijikawa:2021hrf}, BHs with mass above $m_{\rm \tiny min}$ are termed ``high mass,'' and those with mass below the mass gap are ``low mass.''
Thus, there are three subpopulations of BBHs: ``low mass - low mass'', ``high mass - low mass'', and ``high mass - high mass'' binaries.
Here we focus on the last subpopulation, containing two high-mass BHs. This is a conservative assumption. Even though including straddling binary mergers (i.e. ``high-mass - low-mass'' systems) would increase the amount of information, it would also entail considering additional and more complex potential contamination from other astrophysical channels. Therefore, our results can be considered upper bounds in terms of estimating uncertainties on $m_\text{\tiny min}$.

According to Ref.~\cite{Hijikawa:2021hrf}, the ``high-mass - high-mass'' BBH components should have very low spin, with an effective inspiral parameter typically smaller than $\sim 0.1$. We therefore simplify our analysis by setting the BH spins to zero, and we do not recover the spin distribution in our population inference. This assumption is again conservative, in the sense that we neglect possible information carried by the spin distribution of the binary components, but we expect it to have a small impact on our results.

\subsection{Pop III injection}

Motivated by the previous discussion, we consider a simplified model that describes a putative population of mergers beyond the upper mass gap of the following form.

The distribution of the primary BH mass $m_1$ is described by a power law model
 \begin{equation}
     p_{m_1}(m_1|\alpha, m _\text{\tiny min}, m_\text{\tiny max}) 
     \propto m_1^{-\alpha}
 \end{equation} 
with $\alpha = 2$ and normalized to unity across the range $ m _\text{\tiny min} \leq m_1 \leq m_\text{\tiny max}$. 
The distribution of mass ratio is also assumed to follow a power law,
\begin{equation}
    p_{m_2}(q|m_1,\beta) \propto q^{\beta},
    \label{eq:pq}
\end{equation}
constrained within the range $m_\text{\tiny min}/m_1 \leq q \leq 1$.

{
\renewcommand{\arraystretch}{1.4}
\setlength{\tabcolsep}{4pt}
\begin{table*}[!t]
\caption{Parameters described the reference Pop III model, as well as the assumed contaminant population.
In the last line, we also report the corresponding prior ranges adopted in the population inference. Masses are expressed in units of $M_\odot$.
}
\begin{tabularx}{2 \columnwidth}{|X|c|c|c|c|c|c|c|c|c|c|}
\hline
\hline
 Model &\multicolumn{7}{c|}{Pop III} &\multicolumn{3}{c|}{Contaminant population}     
\\
\hline
\hline
$\boldsymbol{\lambda}$ & 
$\alpha$ & 
$\beta$ &  
$m_\text{\tiny min}$ & 
$m_\text{\tiny max}$ &
$\aIII$ &
$\bIII$ &
$\zIII$ &
$\alpha_c$ & 
$\beta_c$ &  
$\gamma_c $
\\
\hline
Injected &
2 & 1 & 130 & 600 & 0.66 & 0.3 & 11.6 &
1 & 0 & 0 
\\
\hline
 Prior & $[-10,10]$ & $[-10,10]$ & $[120,200]$ & $[200, 700]$ &
 $[-10,10]$ & $[-10,10]$ & $[-10,10]$ &
 $[-10,10]$ & $[-10,10]$ & $[-10,10]$ 
 \\
\hline
\hline
\end{tabularx}
\label{Tb:parameters}
\end{table*}
}

The normalized rate density of Pop~III mergers is given by
\begin{align}\label{eq:pop3}
p_z(z|\aIII,\bIII,\zIII) \equiv  \frac{e^{\aIII(z-\zIII)}}{\bIII+\aIII e^{(\aIII+\bIII)(z-\zIII)}},
\end{align}
where $\aIII$, $\bIII$, and $\zIII$ characterize the upward slope at~$z<\zIII$, the downward slope at~$z>\zIII$, and the peak location of the volumetric merger rate density, respectively. Following Ref.~\cite{Ng:2020qpk} , we set $a_{\rm III} = 0.66$, $b_{\rm III}= 0.3$, and $z_{\rm III} = 11.6$.
This leads to a merger rate peak at $z\sim 11.6$ and to a non-negligible merger rate above $z\gtrsim 15$, consistent with theoretical predictions~\cite{Kinugawa:2020, Belczynski:2017, Kinugawa:2020ego, Hijikawa:2021hrf, Liu:2021, Santoliquido:2023}.
We thus write the differential Pop~III BBH merger rate density as
\begin{align}\label{eq:Rate_pop}
\frac{\d  R}{\d m_1 \d m_2}
\propto & \,
p_{m_1} (m_1|\alpha,m_\text{\tiny min},m_\text{\tiny max})
\nonumber 
\\
\times & \,
p_{m_2} (m_2|m_1,\beta)\,
p_{z}(z|\aIII,\bIII,\zIII).
\end{align}
In Fig.~\ref{fig:pop_errors} (top panel), we show the probability distributions of relevant binary parameters (i.e., $m_1$, $q$, and $z$) for both the {\it intrinsic} and {\it observed} populations
(see Sec.~\ref{sec:selectionbias} for details on the selection bias).
We assume an isotropic distribution of binaries in the sky and an isotropic distribution of their orientation.
In Table~\ref{Tb:parameters} we summarize the injected population parameters.

With these definitions, the rate of Pop~III merger events at a given redshift $\d N_\text{\tiny yr}(z )/\d z$ is computed by multiplying the merger rate density by the differential comoving volume and redshift factor:
\begin{equation}
\frac{\d N_\text{\tiny yr}(z )}{\d z}=
R_\text{\tiny pk}
\frac{1}{1+z} \frac{\d V}{\d z } 
p_z(z \mid \aIII, \bIII, \zIII),
\end{equation}
where $R_{\rm pk}$ is the merger rate at the peak of the rate density.
Assuming a rate at the peak normalized in such a way that $R_{\rm pk}=1\,\rm Gyr^{-3}\, yr^{-1}$, the rate of Pop~III mergers in the redshift window $z\in[5,20]$ is $N_\text{\tiny yr}=128\,\rm yr^{-1}$. 

We assume a network of next-generation GW detectors composed of one Einstein Telescope, with triangular configuration and ET-D sensitivity curve, located in Sardinia, Italy, and two Cosmic Explorer interferometers (with 40 km and 20km arm length, respectively) located in the USA (Idaho and New Mexico)~\cite{Iacovelli:2022bbs}.
This network is only able to detect a fraction $\rho \equiv N_{\rm det}/N=0.52$ of all mergers for the reference population introduced above (see Sec.~\ref{sec:selectionbias} below).
Therefore, $N=100, 200$, and $500$  detections in an observation period of $T_\text{\tiny obs} = 1 {\rm yr}$ correspond to a peak rate density of $R_{\rm pk}=1.5, 3.0,\ {\rm and}\ 7.5\,\rm Gpc^{-3}\, yr^{-1}$, respectively.

\subsection{Contamination beyond the upper mass gap}
\label{Sec:Contamination_of_the_upper_mass_gap}

As discussed in the introduction, various possible contamination mechanisms may produce mergers within and above the mass gap. 
In this work, we remain agnostic about these possibilities and parametrically describe the contaminant population with simple power-law models. As we are interested in measuring the location of $m_\text{\tiny min}$, the most important contamination comes from mergers with component masses comparable to $m_\text{\tiny min}$, whose impact is controlled, in practice, by the assumed rate. 

The contaminant population is chosen to be characterized by a power law scaling of the form \eqref{eq:Rate_pop} with $\alpha_c = 1$, $\beta_c = 0$, $m_{\text{\tiny min},c} = 120 M_\odot$,
$m_{\text{\tiny max},c} = 700 M_\odot$ and 
a flat merger rate density $\d R/\d z \propto (1+z)^{\gamma_c}$ with $\gamma_c = 0$.
Notice that we purposely limit attention to mergers from the contaminant population close to the edge of the upper mass gap, i.e., we assume $m_{\text{\tiny min},c}$ to be close to the feature in the Pop III mass distribution. 
If a sufficiently precise model for the contaminant population were available, more information from smaller masses may help us to constrain the contaminant population, and thus reduce the uncertainties on $m_\text{min}$, making our choice conservative. This assumption corresponds to restricting the analysis to the range of primary masses of interest, and thus we do not include $m_{\text{\tiny min},c}$ and $m_{\text{\tiny max},c}$ in the population inference. 

Most importantly, we vary the overall merger rate of the contaminant population to chart how this would affect the uncertainty on $m_\text{\tiny min}$.
For simplicity, we define the number of detectable events of the contaminant population to be $N_\text{\tiny det}^\text{\tiny C}$.
In practice, we fix the ratio 
\begin{equation}
    f_\text{\tiny det} = \frac{N_\text{\tiny det}^\text{\tiny C}}{N_\text{\tiny det}}.
\end{equation}
For the assumed population, the detection fraction is $\rho \equiv N_\text{\tiny det}/N = 0.58$. 
This fraction is very similar to the one obtained for the reference Pop III model, so $f_\text{\tiny det}$  roughly corresponds to the ratio between the {\it intrinsic} number of mergers for both the Pop III and contaminant populations.

\begin{figure*}
\centering
\includegraphics[width=0.49 \textwidth]{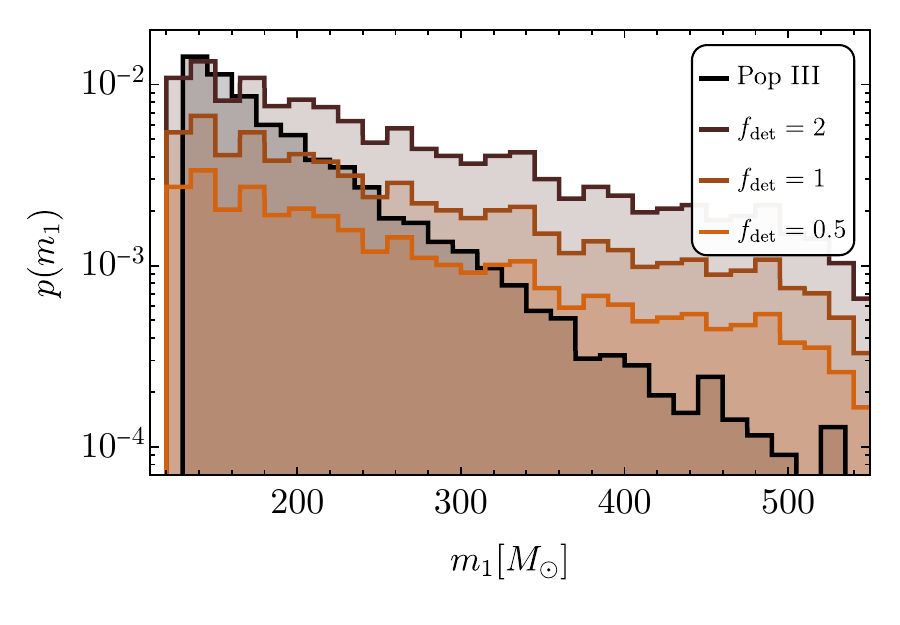}
\includegraphics[width=0.49 \textwidth]{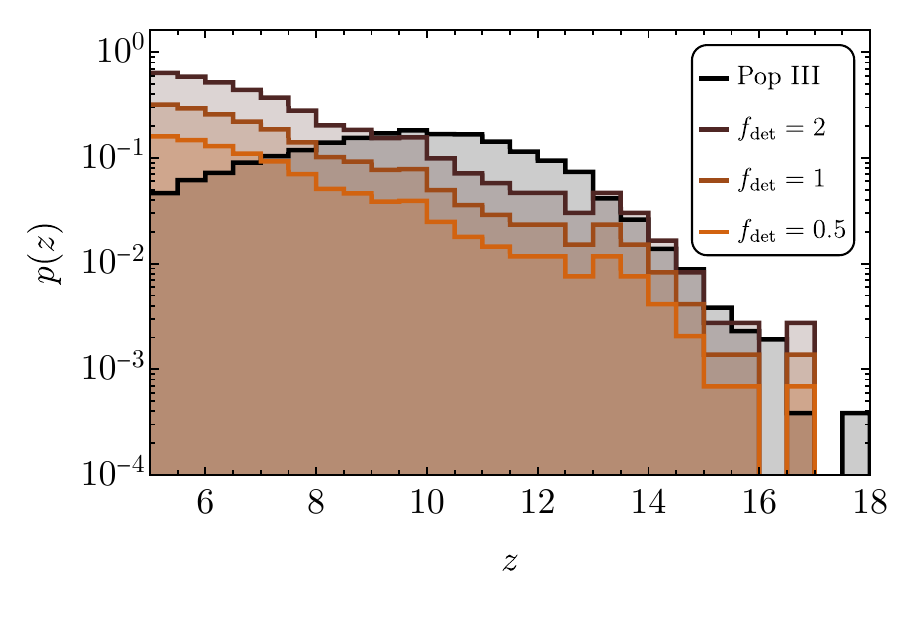}
\caption{
The plot of the {\it observed} distribution of primary mass (left) and redshift (right) for both Pop III and contaminant populations, normalized to different relative fractions $f_\text{\tiny det}$ as indicated in the inset. 
}
\label{fig:injection12_pop}
\end{figure*}

In Fig.~\ref{fig:injection12_pop} we show the {\it observed} distribution (i.e., after selection effects are included: see Sec.~\ref{sec:selectionbias}) of the primary mass and redshift, for both the Pop III and the contamination channels. 
As one can see from the figure, assuming $f_\text{\tiny det} = 2$ would provide a comparable rate of events in the mass range close to $m_\text{\tiny min}$. Larger rates would completely erase the gap's edge.
Also, due to the shallower tilt of $p(m_1)$ for the contaminant, a sizeable contamination of this sort is expected to completely erase the sensitivity to the high-mass tail of Pop III mergers.
Also, the contaminant population with $f_\text{\tiny det} > 0.1$ dominates the merger rate in the lowest redshift bins considered in this work, where most precise measurement of masses are achieved: see also Fig.~\ref{fig:pop_errors}. 

\section{Methods}
\label{Sec:Methods}

In this section we describe the methods we adopt to simulate future observations by next-generation detectors, including parameter estimation and Bayesian population analysis. The knowledgeable reader can directly proceed to the next section, where we report the results of our analysis.  

\subsection{Fisher information matrix forecasts}

We estimate the measurability of source properties using the Fisher information matrix (FIM) approach, as is typically done to assess the parameter estimation capabilities of next-generation GW detectors when dealing with large injection campaigns~\cite{PhysRevD.46.5236,PhysRevD.47.2198,Cutler:1994ys,Poisson:1995ef,Berti:2004bd,Ajith:2009fz,Cardoso:2017cfl}  (see~\cite{Vallisneri:2007ev,Rodriguez:2013mla} for the potential limitations of this approach.)
In this work we use {\tt GWFast}~\cite{Iacovelli:2022mbg,Iacovelli:2022bbs}, a numerical package to evaluate SNRs and carry out FIM forecasts for measurement uncertainties with a network of next-generation detectors.
Other parameter estimation codes have been recently  developed, including \textsc{GWBENCH}~\cite{Borhanian:2020ypi,Borhanian:2022czq};   \textsc{GWFISH}~\cite{Dupletsa:2022scg}; \textsc{TiDoFM}
\cite{Chan:2018csa,Li:2021mbo};
and the code used in Ref.~\cite{Pieroni:2022bbh}.
Results from these codes are consistent with each other~\cite{Branchesi:2023mws, Iacovelli:2022bbs}.
Our approach is complementary to Ref.~\cite{Fairhurst:2023beb}, which considered Bayesian parameter estimation of simulated distant BBHs beyond the mass gap with a similar network of next-generation detectors. 
Despite using different detector designs and parameter estimation methods, we found qualitative agreement with their results (see Appendix~\ref{app:PEexamples}).

The measured output $s(t)$ of a GW observatory is the sum of the GW signal $h (t, {\boldsymbol \theta})$ and the detector noise $n(t)$, here assumed to be Gaussian and stationary with zero mean. The posterior distribution for the hyperparameters ${\boldsymbol \theta}$ can be approximated by
\be\label{pos_dist_F}
p ({\boldsymbol \theta}| s)
\propto \pi ({\boldsymbol \theta}) 
\exp
\left[-\frac{1}{2}(h ({\boldsymbol \theta}) - s|h ({\boldsymbol \theta}) - s)
\right]\,,
\ee
where $\pi ({\boldsymbol \theta})$ indicates the prior distribution and we have defined the inner product
\be\label{innprod}
(g|h) = 4\Re\int_{f_\text{\tiny min}}^{f_\text{\tiny max}} \frac{\tilde{g}^* (f) \tilde{h} (f) }{S_n(f)} \,\dd{f}\,.
\ee
In Eq.~\eqref{innprod}, the tilde denotes a Fourier transform, $S_n(f)$ is the detector noise power spectral density (PSD), while $f_\text{\tiny min}$ and $f_\text{\tiny max}$ are the detector minimum and maximum frequency of integration, respectively.
The SNR is defined as ${\rm SNR}\equiv\sqrt{(h|h)}$.
More details on the detector noise PSD used here can be found in Ref.~\cite{Iacovelli:2022bbs}.
 
In the limit of large SNR, one can Taylor expand Eq.~\eqref{pos_dist_F} and get (focusing only on statistical uncertainty and neglecting the noise realization dependence)
\be\label{eq:FisherPost}
p ({\boldsymbol \theta}| s)
\propto \pi ({\boldsymbol \theta}) 
\exp\left[-\frac{1}{2}\Gamma_{ab} \Delta {\boldsymbol \theta}^a \Delta {\boldsymbol \theta}^b\right]\,,
\ee
where $\Delta {\boldsymbol \theta} = {\boldsymbol \theta}_\text{\tiny p} - {\boldsymbol \theta}$; ${\boldsymbol \theta}_\text{\tiny p}$ are the posterior mean values that coincide, by construction, with the true binary parameters ${\boldsymbol \theta}_p={\boldsymbol \theta}_\text{\tiny true}$, and the Fisher matrix is defined as
\be
\Gamma_{ab} \equiv \Big( \frac{\partial h}{\partial {\boldsymbol \theta}^a} \bigg | 
\frac{\partial h}{\partial {\boldsymbol \theta}^b} \Big)_{{\boldsymbol {\boldsymbol \theta}} = {\boldsymbol \theta}_\text{\tiny p}}\,.
\ee
The errors on the parameters are then given by $\sigma_a = \sqrt{\Sigma^{aa}}$, where $\Sigma^{ab} = (\Gamma^{-1})^{ab}$ is the covariance matrix.

We consider BBHs on quasicircular orbits, which are described by 15 parameters
\begin{equation}\label{eq:paramBBH}
{\boldsymbol \theta}
= \{{\cal M}_c, \eta, d_L, \theta, \phi, \iota, \psi, t_c, \Phi_c, \chi_{1,j}, \chi_{2,j} \}\,,    
\end{equation}
where $j = \{x,y,z\}$ (see e.g.~\cite{Maggiore:2007ulw}).
Here  ${\cal M}_c$ denotes the detector-frame chirp mass; $\eta$ the symmetric mass ratio; $d_L$ the luminosity distance to the source; $\theta$ and $\phi$ the sky position coordinates, defined as $\theta=\pi/2-\delta$ and $\phi$ (with $\phi$ and $\delta$ right ascension and declination, respectively); $\iota$ the inclination angle of the binary with respect to the line of sight; $\psi$ the polarization angle; $t_c$ the time of coalescence; $\Phi_c$ the phase at coalescence; and $\chi_{i,j}$ the dimensionless spin of the object $i=\{1,2\}$ along the axis $j = \{x,y,z\}$.

We use the inspiral–merger–ringdown (IMR) phenomenological  waveform model IMRPhenomHM, which includes the contribution of the higher–order harmonics $(\ell, m) = (2, 1), (3, 2), (3, 3), (4, 3)$, and $(4, 4)$ in addition to the dominant $(2, 2)$ multipole~\cite{London:2017bcn,Kalaghatgi:2019log}. 
This waveform neglects precession effects, expected to be subdominant for the slowly spinning Pop III mergers we consider here.

We finally translate the posterior distributions on the chirp mass and symmetric mass ratio into (ordered) component masses $(m_1,m_2)$ (including the associated Jacobian factor), and determine the source redshift posterior assuming a standard $\Lambda$CDM cosmology~\cite{Planck:2018vyg}. 

\subsection{Hierarchical Bayesian inference}
\label{Sec:Hierarchical_Bayesian_analysis}

Our statistical analysis for the recovery of the hyperparameters, ${\boldsymbol \lambda}$, of the Pop III model is based on the hierarchical Bayesian framework, with the inclusion of selection effects and measurement uncertainties.

The posterior probability distribution of ${\boldsymbol \lambda}$ is computed adopting the likelihood~\cite{Mandel:2018mve}
\begin{align}
\frac{p({\boldsymbol \lambda}|{\bm d})}{\pi({\boldsymbol \lambda}) }
 \propto 
e^{- N \rho({\boldsymbol \lambda})}
\prod_{i=1}^{N_\text{\tiny obs}}
\frac{1}{{\cal S}_i}\sum_{j=1}^{{\cal S}_i} 
\frac{N p_\text{\tiny pop}(^j{\boldsymbol \theta}_i|{\boldsymbol \lambda})}{\pi(^j{\boldsymbol \theta}_i)}.
\label{eq:populationPosterior_discrete}
\end{align}
We take uniform priors $\pi({\boldsymbol \lambda})$ on the domains of each parameter, as defined in Table~\ref{Tb:parameters}. 
The index $i$ labels each observed GW event, $j$ labels the points in the posterior distribution of each of its parameters~\eqref{eq:paramBBH}, and
${\cal S}_i$  identifies the number of samples adopted to compute the Monte Carlo integration over the posterior of each event.
We marginalize over number of mergers (or overall population rates) with log-uniform priors.
We sample the posterior on the hyperparameters ${\boldsymbol \lambda}$ using the {\tt emcee} sampler~\cite{Foreman-Mackey:2012any}.

\subsection{Selection bias}\label{sec:selectionbias}		

We compute the selection bias using a threshold on the detector network SNR, defined as 
\begin{equation}
\text{SNR}^2 = \sum_{i} \text{SNR}_{i}^2,
\end{equation}
where the index $i$ identifies the three next-generation detectors assumed in this work. 
As usually assumed in the literature, we set the detection threshold to be a network SNR = 12.
While it is possible to introduce more sophisticated detection statistics (such as the false alarm rate, or the probability ``p-astro'' of events being of astrophysical origin), we expect these simplified conditions to be sufficient for our present purposes. 

The bias factor, or detection efficiency, $\rho \equiv N_\text{\tiny det}/N$ is approximately calculated by Monte Carlo integration, as usual in LVK analyses (see e.g.~\cite{LIGOScientific:2021psn}).
For this purpose, we compute the SNR for an arbitrarily injected population that covers sufficiently densely the BBH intrinsic parameter space.
We recover $N_\text{\tiny found}$ BBHs with above-threshold SNR in our injection campaign, which is used to chart the observable parameter space of the detector network. 

We evaluate the selection fraction by reweighting the population with
hyperparameters ${\boldsymbol \lambda}$ as~\cite{Mandel:2018mve}
\begin{equation}
\rho({\boldsymbol \lambda}) = 
\frac{1}{N_\text{\tiny inj}} 
\sum_{j=1} ^{N_\text{\tiny found}}
\frac{p_\text{\tiny pop}({{\boldsymbol \theta}_j}|{\boldsymbol \lambda})}
 {p_\text{\tiny inj}({{\boldsymbol \theta}_j})},
    \label{eq:sel-effects}
\end{equation}
where $N_\text{\tiny found}$ is the number of recovered events, $N_\text{\tiny inj}$ is the total number of injections (including the low-SNR, unobservable ones), and $p_\text{\tiny inj}(\boldsymbol{\theta})$ is the reference distribution from which injections were built.
The detailed properties of the reference population $p_\text{\tiny inj}(\boldsymbol{\theta})$ used to estimate $\rho(\boldsymbol{\lambda})$ are irrelevant, as they factorize out in Eq.~\eqref{eq:sel-effects}.

\section{Results}	\label{Sec:Results}	

To take into account uncertainties in the merger rate of both the Pop III channel and the supposed contaminant population, we simulate different scenarios by varying the number of detections in both sectors. 
In practice, we assume several Pop III detections $N_\text{\tiny det} = \{50, 100,200,500\}$ in one-year observations with next-generation detectors, and a contamination fraction $f_\text{\tiny det} = \{0,0.1,0.5,1,2\}$.

\begin{figure}[t!]
\centering
\includegraphics[width=0.49 \textwidth]{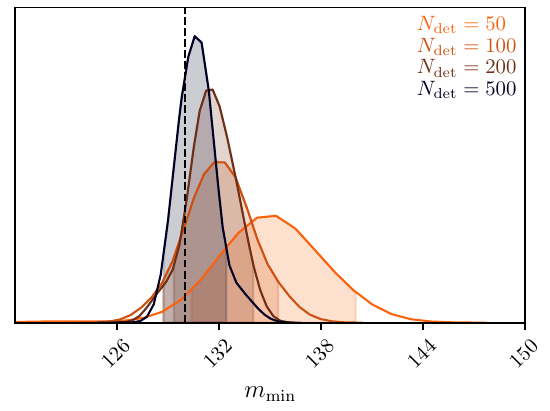}
\label{fig:injection1}
\caption{
Posterior distribution of $m_\text{\tiny min}$ assuming $N_\text{\tiny det} = 50$,  $100$, $200$ and $500$ detections and negligible contamination.
The vertical black line is the injected value of $m_\text{\tiny min} = 130 M_\odot$, while the colored vertical bands bracket the 90\% C.L. range.}
\end{figure}

In Fig.~\ref{fig:injection1} we show the posterior distribution of $m_\text{\tiny min}$ for different values of $N_\text{\tiny det}$ and negligible contamination. 
The injected values are always recovered within the $2\sigma$ confidence level, but with a systematic shift towards values of $m_\text{\tiny min}$ that are slightly larger than the injected value. This is due to the asymmetry of $p_{m_1}$ around $m_\text{\tiny min}$. 
As a function of the assumed number of detections, the uncertainty at 90\% C.L. is well approximated by
\begin{equation}
\frac{\Delta {m_\text{\tiny min}}}{{m_\text{\tiny min}}} 
\simeq 
0.04 
\left( \frac{N_\text{\tiny det}}{100} \right )^{-1/2}\,.
\end{equation}

\begin{figure*}\label{fig:pos1}
\centering
\includegraphics[width=0.49 \textwidth]{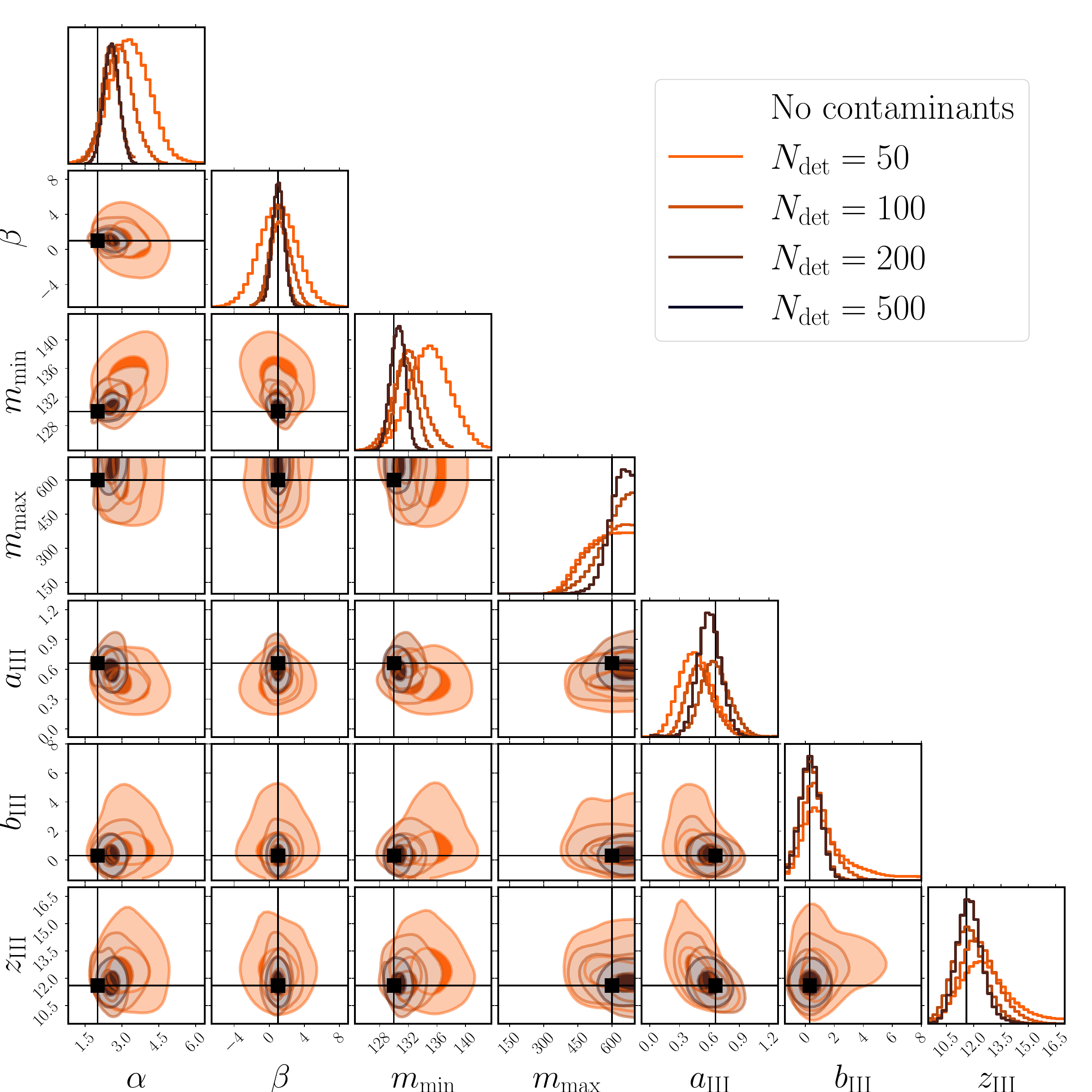}
\includegraphics[width=0.49 \textwidth, trim=0 -0.85cm 0 0 ]{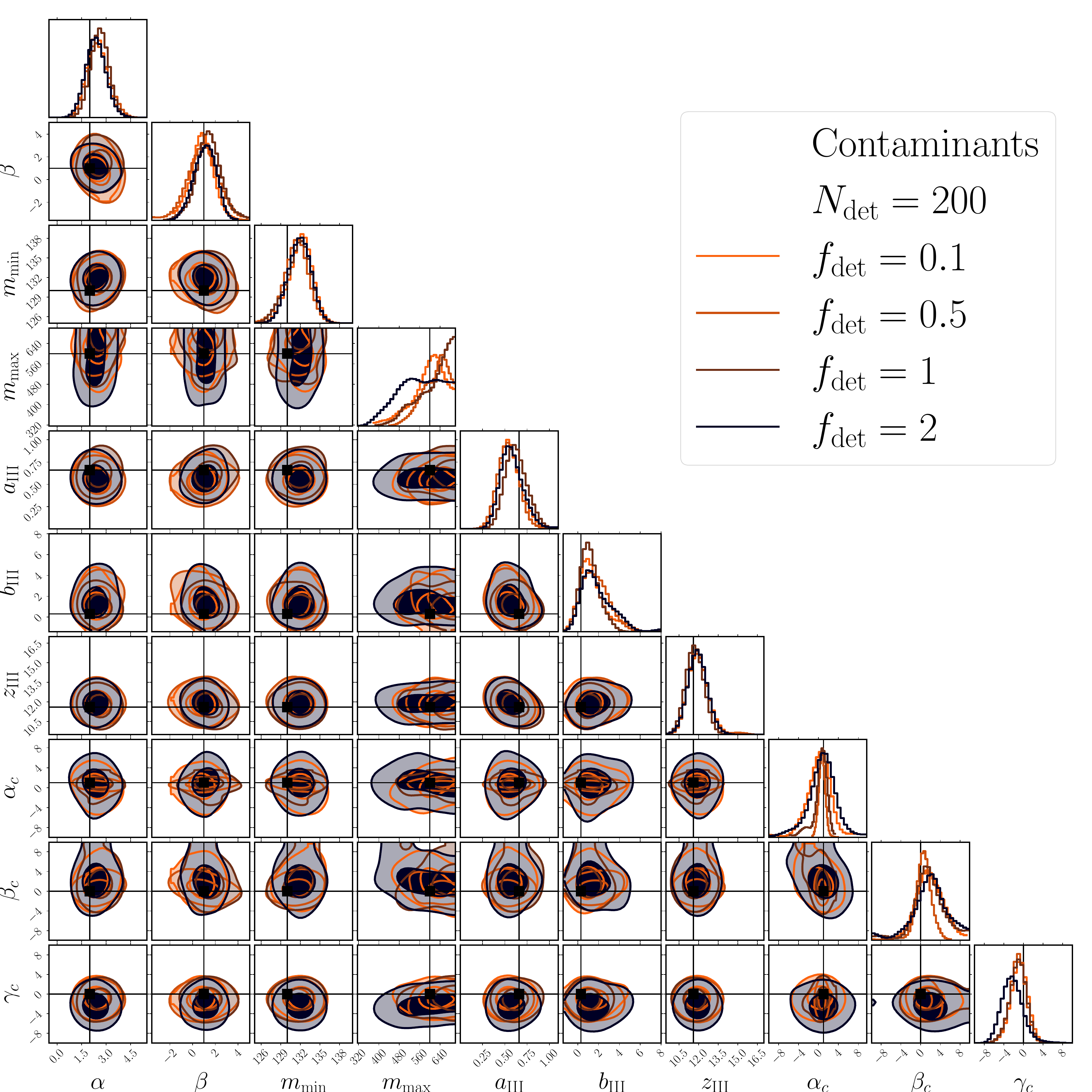}
\caption{
Left panel:
same as Fig.~\ref{fig:injection1}, but now including posteriors for all of the population parameters. 
Right panel:
posterior distributions for both the Pop III and contaminant population parameters, for $N_\text{\tiny det} = 200 $ and four different values of $f_\text{\tiny det}$.}
\end{figure*}

In Fig.~\ref{fig:pos1} (left panel) we report the full posterior distribution.
We observe a positive correlation between the primary mass power law index $\alpha$ and $m_\text{\tiny min}$. This is because, given the observed BBH population, assuming a larger value of $m_\text{\tiny min}$ shrinks the $m_1$ domain, forcing the tilt to become steeper. 
As a consequence of the asymmetry of the population, the tilt is biased towards larger values (which are anyway compatible with the injection within the $2\sigma$ C.L.). 
We do not observe strong degeneracies between $m_\text{\tiny min}$ and any other population parameters. 
As anticipated in the discussion of Fig.~\ref{fig:injection12_pop}, given the relatively steep $m_1$ distribution, measurements of the large mass cut-off $m_\text{\tiny max}$ are not possible: at best we can set lower bounds, which become tighter as $N_\text{\tiny det}$ grows.

\begin{figure}[t]
\centering
\includegraphics[width=0.49\textwidth]{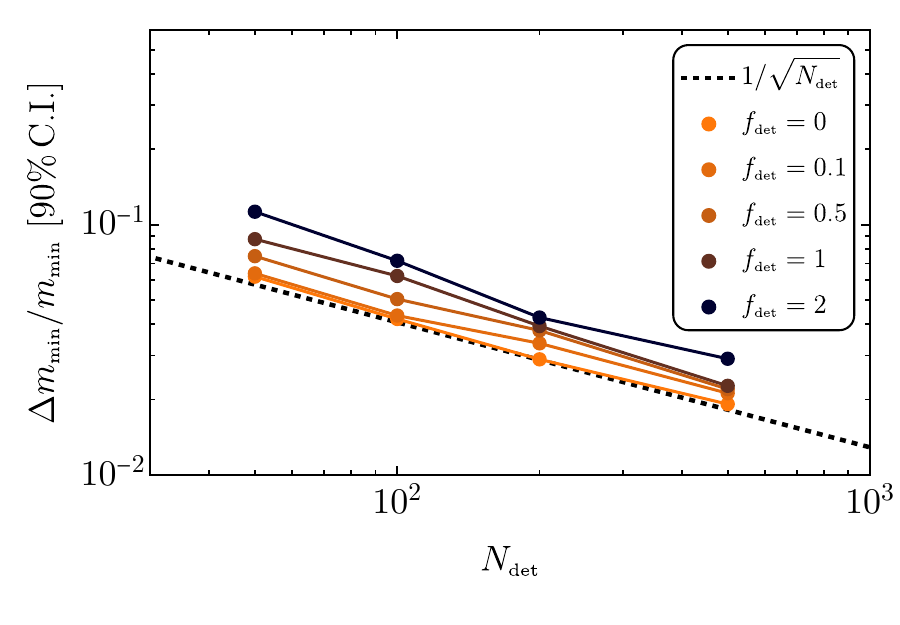}
\caption{Scaling of the relative error (at 90\% C.I) as a function of the number of detections, compared to the expected scaling of statistical uncertainty $\approx 1/\sqrt{N_\text{\tiny det}}$ to guide the eye.
}
\label{fig:moneyplot}
\end{figure}

We now evaluate the relevance of a supposed contaminant population by injecting populations with increasing values of $f_\text{\tiny det}$. The results are shown in the right panel of Fig.~\ref{fig:pos1} and in Fig.~\ref{fig:moneyplot}, where we report the 90 \% C.L. uncertainties on $m_\text{\tiny min}$ as we vary both $N_\text{\tiny det}$ and $f_\text{\tiny det}$.

The results in Fig.~\ref{fig:moneyplot} indicate that including a sizeable contaminant population degrades the sensitivity to $m_\text{\tiny min}$ by a relatively small amount. 
There are two competing trends in this case. A larger contamination fraction means that a larger number of contaminant events populate the mass range close to $m_\text{\tiny min}$, potentially erasing the feature we would like to observe. 
On the other hand, more observations of the contaminant subpopulation lead to better constraints and to smaller degeneracies between population parameters.
The combination of these two trends leads to a mild dependence of $\Delta m_\text{\tiny min}/m_\text{\tiny min}$ on $f_\text{\tiny det}$.
A good fit to our results is  
\begin{align}
    \Delta m_\text{\tiny min}/m_\text{\tiny min} \propto 0.025 \cdot f_\text{\tiny det}
    \quad &\text{for} \quad 
    N_\text{\tiny det} = 50,
    \nonumber 
    \\
   \Delta m_\text{\tiny min}/m_\text{\tiny min} \propto 0.005 \cdot f_\text{\tiny det}
   \quad &\text{for} \quad 
   N_\text{\tiny det} = 500.
\end{align}

In the right panel of Fig.~\ref{fig:pos1} we report the full posterior distribution of the analysis of a synthetic catalog containing both Pop III binaries and the contaminant population. Here we fix $N_\text{\tiny det} = 200$ and vary the fraction $f_\text{\tiny det}$. 
Overall we observe small changes in the Pop III population parameter posteriors, due to the relatively mild effect of the contaminant. 
The largest effect is a weakening of the lower bound on $m_\text{\tiny max}$ for large values of $f_\text{\tiny det}$, as anticipated in the previous section (see also Fig.~\ref{fig:injection12_pop}).
In addition, the presence of a contaminant population with a redshift-independent merger rate density reduces the accuracy with which we can determine the high-redshift slope of the rate density, $b_\text{III}$.

\section{Astrophysical implications and conclusions}\label{sec:astroimpl}

We now wish to quantify how the uncertainties in $m_\text{\tiny min}$ translate into uncertainties in the physical parameters controlling the physics of the mass gap, and in particular (following the discussion in 
Ref.~\cite{2020ApJ...902L..36F}) how they translate into uncertainties in the nuclear reaction rate \crate{}.

Within the isolated binary evolution paradigm, this reaction rate is considered to be the dominant unknown physical parameter controlling the measured value of $m_\text{\tiny min}$~\cite{Farmer:2019jed,Mehta:2021fgz}.
Inferring this rate is difficult due to the negligible cross-section of this reaction at temperatures relevant for He burning in stars~\cite{An:2015xma}.
Current nuclear experiments can only explore higher energy regimes, from which the relevant rates are extrapolated to smaller (astrophysically relevant) temperatures. 
Because of the complex energy dependence of the cross-section, this extrapolation leads to large uncertainties~\cite{deBoer:2017ldl,Friscic:2019eow} (we neglect the possible effect of physics beyond the Standard Model on the location of the mass gap: see e.g.~\cite{Sakstein:2020axg}).

Following Ref.~\cite{2020ApJ...902L..36F}, we parametrize the relation between $m_\text{\tiny min}$ and \crate{} in terms of number of standard deviations $\sigma_{\rm C12}$ from the median rate given in \texttt{STARLIB}~\cite{Sallaska:2013xqa}. Each value of $\sigma_{\rm C12}$ thus corresponds to a different rate \crate{}. We fit the relation between $m_\text{\tiny min}$ and $\sigma_{\rm C12}$ around the mean measured value reported in Ref.~\cite{2020ApJ...902L..36F}. This relation can be written as 
\begin{equation}
    m_\text{\tiny min}/M_\odot \simeq 
    130
     -9.4 \sigma_{\rm C12} 
     + 1.8  \sigma_{\rm C12} ^2\,.
\end{equation}
Assuming $N_\text{\tiny det} = 50$ with no significant contamination, measurements of $m_\text{\tiny min}$ would result in a reduction of the relative uncertainty on \crate{} to about $\approx 26\%$ of its current value. This can be further reduced to $\approx 8 \%$ with $N_\text{\tiny det} = 500$.
For large contamination ($f_\text{\tiny det} = 2$), the relative uncertainty would be limited to $\approx 53\%$ for $N_\text{\tiny det} = 50$ and $\approx 12\%$ for $N_\text{\tiny det} = 500$, respectively.

It is important to stress that pinning down the location of the {\em upper} edge of the mass gap may also help with the interpretation of events close to the {\em lower} edge. Even though uncertainties remain, the {\em width} of the upper mass gap stays relatively constant with respect to variations of $\sigma_{\rm C12}$, and it is reported to be $83^{+5}_{-8} M_\odot$ in Ref.~\cite{Farmer:2019jed}.
Therefore, constraints on the upper end of the gap could translate directly into constraints on the lower end.

Future observations of the upper BH mass gap will further aid in providing important astrophysical implications for BH formation. We highlight two examples.

The first example is the interpretation of the ${35 M_\odot}$ peak in the mass distribution as formed from a PPISN pile-up.
Current data from the LIGO-Virgo-KAGRA GWTC-3 catalog show a preference for a peak in the mass distribution around $35M_\odot$~\cite{Farah:2023vsc, LIGOScientific:2021psn,Sadiq:2021fin,Callister:2023tgi}.
This BH pile-up could be naturally produced by the PPISN mechanism, which reduces the mass of the heaviest stars to a value that results in similar-mass BHs \citep{Woosley:2017,2020A&A...640A..56R}.
However, according to recent studies~\cite{Golomb:2023vxm,Hendriks:2023yrw}, the observed feature appears at too light masses, and it is unlikely to come from PPISN physics due to the unreasonably large inferred nuclear reaction rates (but see~\cite{Ghodla:2023ymi}).
The large expected contamination from astrophysical channels discussed in the introduction further jeopardizes our ability to reach solid conclusions on the location of the lower edge PISN gap using only observations of low-mass BBH mergers. On the contrary, observations of events in the ``far side'' could shed some light on this feature. 

The second example is the interpretation of events falling in the mass gap.
Currently, the interpretation of events in the mass gap such as GW190521~\cite{LIGOScientific:2020iuh} is challenging, due to uncertainties on the actual location of the gap within the BH mass spectrum. Clean observations of the upper edge could potentially confirm the need for alternative explanations for GW190521. Some of the proposed scenarios include a straddling binary~\cite{Nitz:2020mga,Fishbach:2020qag}, 
a second (or higher) generation merger~\cite{Rodriguez:2016kxx,Gerosa:2017kvu,Gerosa:2021mno}, super-Eddington accretion~\cite{vanSon:2020zbk,Safarzadeh:2020vbv,Cruz-Osorio:2021qbr}, primordial black holes~\cite{DeLuca:2020sae,Clesse:2020ghq,Franciolini:2022tfm}, or new physics~\cite{Straight:2020zke,Sakstein:2020axg,CalderonBustillo:2020fyi}.

To conclude, the better sensitivity of next-generation GW detectors at frequencies as low as a few Hz allows us to probe BH masses above $100M_\odot$ at high redshift. The observation of such events would allow us to investigate binary mergers originating from metal-poor Pop~III stars. The measurement of a sharp increase (or a bump) in the primary mass distribution in the range $\sim100$--$150M_\odot$ could indicate
the existence of a population of isolated binaries with component masses on the ``far side'' beyond the upper edge of the mass gap.
In this paper we have estimated the accuracy with which the location of the upper end of the mass gap could be measured with a network of next-generation detectors, and we have shown that they would allow us to place tight constraints on the location of the gap $m_\text{\tiny min}$ and on the physical parameters that control $m_\text{\tiny min}$, such as the \crate{} reaction rate.

\begin{figure*}
\centering
\includegraphics[width=0.49 \textwidth]{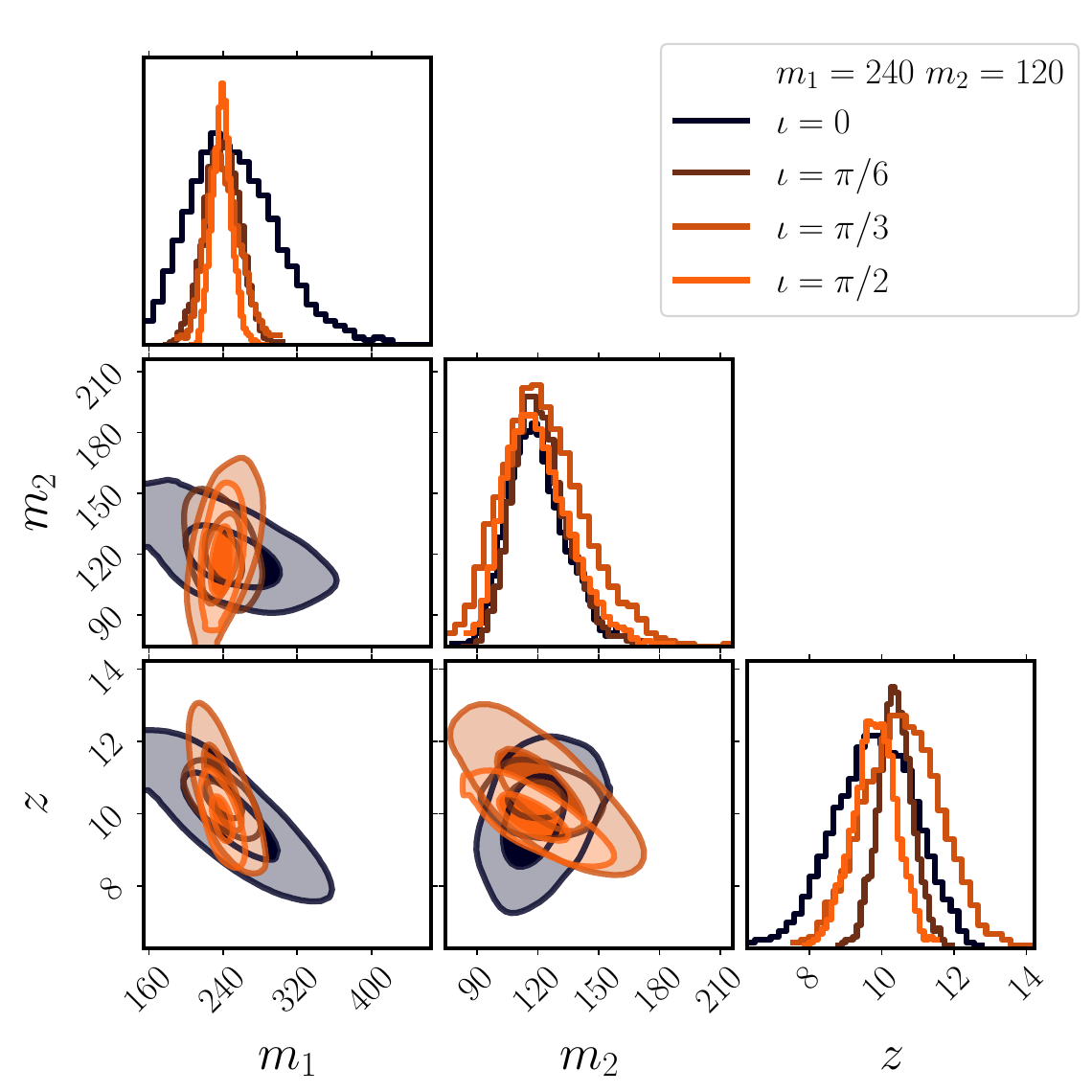}
\includegraphics[width=0.49 \textwidth]{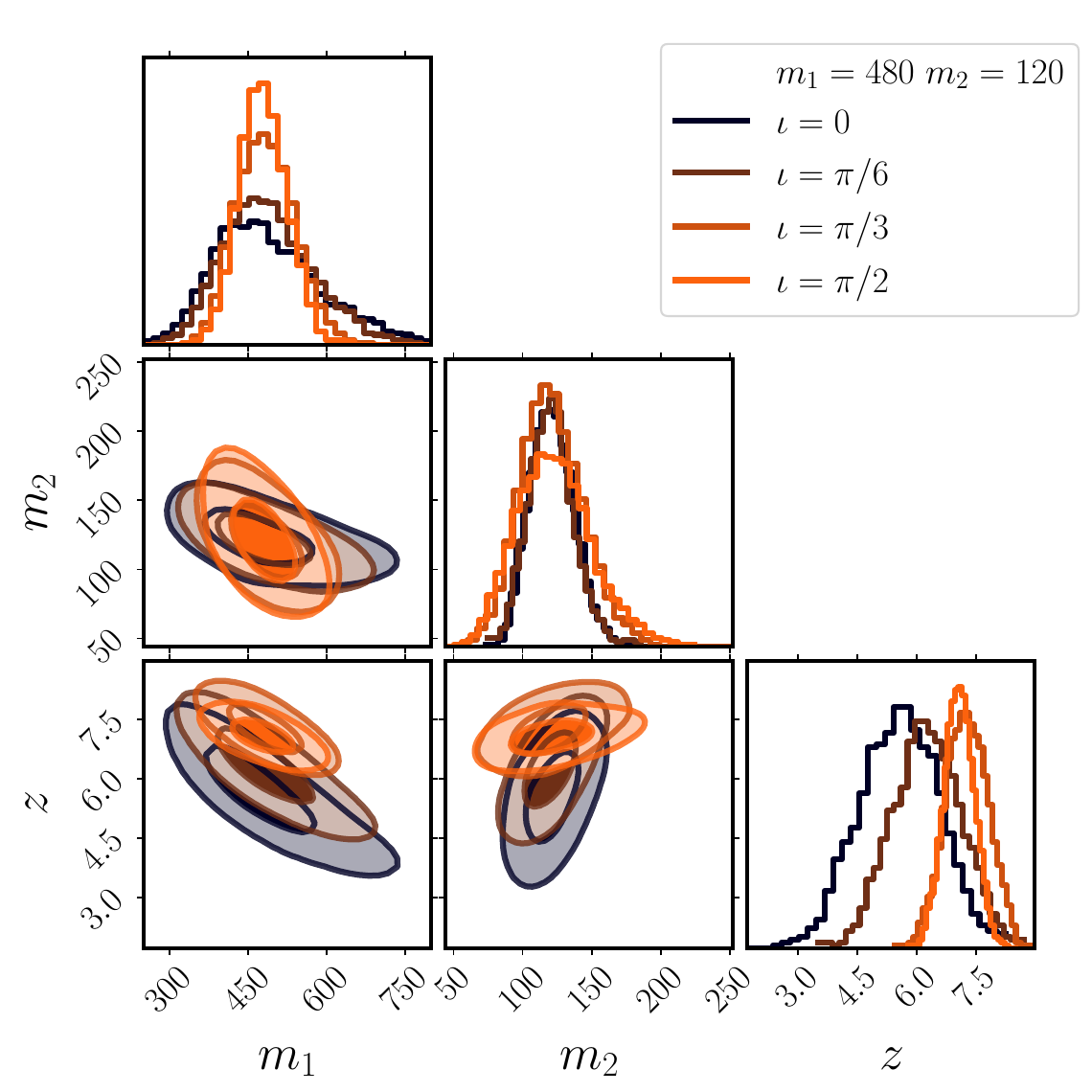}
\caption{
The posterior distribution of masses and redshift for two binary systems at a distance resulting in SNR = 30, and for different binary inclinations $\iota$ (as indicated in the legend). 
Left panel:
$(m_1,m_2) = (240 M_\odot, 120 M_\odot)$.
Right panel:
$(m_1,m_2) = (480 M_\odot, 120 M_\odot)$.
}
\label{fig:pos1PE}
\end{figure*}

\vspace{.3cm}
\acknowledgments
\vspace{-.1cm}
We dedicate this paper to Chris Belczynski, who always pushed himself and others beyond the far side.
We thank R.~Cotesta for collaboration in the early stages of this work, and F.~Santoliquido for insightful comments and suggestions. 
E.B., K.K. and L.R. are upported by NSF Grants No. AST-2006538, PHY-2207502, PHY-090003 and PHY-20043, by NASA Grants No. 20-LPS20-0011 and 21-ATP21-0010, by the John Templeton Foundation Grant 62840, by the Simons Foundation, and by the Italian Ministry of Foreign Affairs and International Cooperation Grant No.~PGR01167.
This work was carried out at the Advanced Research Computing at Hopkins (ARCH) core facility (\url{rockfish.jhu.edu}), which is supported by the NSF Grant No.~OAC-1920103.
K.K. is supported by the Onassis Foundation - Scholarship ID: F ZT 041-1/2023-2024.
The following software libraries were used at various stages in the analysis for this work, in addition to the packages explicitly mentioned above: \software{numpy}~\cite{vanderWalt:2011bqk}, {\tt matplotlib}~\cite{Hunter:2007ouj}, \software{scipy}~\cite{2020SciPy-NMeth}, {\tt astropy}~\cite{Astropy:2013muo}, \software{filltex}~\cite{2017JOSS....2..222G}.
The authors also acknowledge the Texas Advanced Computing Center (TACC) at The University of Texas at Austin for providing HPC resources that have contributed to the research results reported within this paper. URL: \url{http://www.tacc.utexas.edu}~\cite{10.1145/3311790.3396656}.

\appendix 

\section{Fisher information matrix-based parameter uncertainties}\label{app:PEexamples}

We report here a few examples of posterior distributions for the most relevant BBH parameters ${\boldsymbol \theta}$.
We focus on the same BBH systems analyzed in Ref.~\cite{Fairhurst:2023beb} in order to compare our simplified FIM framework to the results of a more exhaustive Bayesian parameter estimation. We find good qualitative agreement with their results. 

In Fig.~\ref{fig:pos1PE} we show the posterior distributions for two BBH systems with masses 
$(m_1,m_2) = (240 M_\odot, 120 M_\odot)$
and 
$(m_1,m_2) = (480 M_\odot, 120 M_\odot)$, 
assumed to be placed at the optimal sky location for the detector network and at a distance corresponding to a network SNR = 30.
Following Ref.~\cite{Fairhurst:2023beb}, and to show the relevance of source inclination in the determination of the luminosity distance, we  consider different binary orientations: $\iota = 0, \pi/6, \pi/3, \pi /2$.
Waveform models containing only the dominant $(2, 2)$ mode suffer from distance-inclination degeneracy, which is especially relevant for nearly face-on ($\iota = 0$) binaries, because the GW amplitude scales as $ (1-\iota^2/2)/d_L$.
This degeneracy is alleviated by including higher-order multipoles in the waveform and by considering a network of detectors rather than a single detector.

Figure~\ref{fig:pos1PE} should be compared with Fig.~9 (posteriors on component masses) and Fig.~10 (posteriors on redshift) of Ref.~\cite{Fairhurst:2023beb}. 
By construction, in the FIM approach the mean value of the binary parameters is the injected value, while we sample the multivariate Gaussian posterior to show uncertainties and parameter correlations. 
The relative uncertainties on the relevant parameters for these moderate-SNR events are close to $20\%$ at $1 \sigma$. The relative uncertainties at the $90\%$ credible intervals agree within a factor of $\lesssim2$ with the results of Ref.~\cite{Fairhurst:2023beb} for all the cases shown here.

\bibliographystyle{apsrev4-1}
\bibliography{refs}

\end{document}